# Are Uranus & Neptune responsible for Solar Grand Minima and Solar Cycle Modulation?

*Sharp G.J.,*
*Melbourne Australia.*
*Email: gs_qad@hotmail.com*

## Abstract

Detailed solar Angular Momentum (AM) graphs produced from the Jet Propulsion Laboratory (JPL) DE405 ephemeris display cyclic perturbations that show a very strong correlation with prior solar activity slowdowns. These same AM perturbations also occur simultaneously with known solar path changes about the Solar System Barycentre (SSB). The AM perturbations can be measured and quantified allowing analysis of past solar cycle modulations along with the 11,500 year solar proxy records (14C & 10Be). The detailed AM information also displays a recurring wave of modulation that aligns very closely with the observed sunspot record since 1650. The AM perturbation and modulation is a direct product of the outer gas giants (Uranus & Neptune). This information gives the opportunity to predict future grand minima along with normal solar cycle strength with some confidence. A proposed mechanical link between solar activity and planetary influence via a discrepancy found in solar/planet AM along with current AM perturbations indicate solar cycle 24 & 25 will be heavily reduced in sunspot activity resembling a similar pattern to solar cycles 5 & 6 during the Dalton Minimum (1790-1830).

## Introduction

Solar system dynamics have been postulated as the main solar driver for many decades. Jose (1965)[4] was the first to associate a recurring solar system pattern of the 4 outer planets (179 years). He suggested this pattern correlates with the modulation of the solar cycle. New research via this study suggests that over the past 6000 years the 179 year cycle cannot be maintained and is closer to a 172 year cycle which aligns with the synodic period of Uranus & Neptune (171.44 years). Later Landscheidt (2003)[5] progressed the planetary influence theories further by associating quasi-cyclic negative torque readings or "zero crossings" (AM readings going below zero) that can occur near grand minima. It has been found since that the negative readings occur in the general region of most grand minima but such records are not a reliable method of predicting the timing and strength of grand minima at the solar cycle level.

Other studies detailed the orbit path of the Sun around the SSB that showed a balanced trefoil pattern during times of "normal" solar cycles. Charvàtovà (2000)[2] shows this pattern or pathway as it moves to a disordered state during times of solar slowdown and are a direct result of the Uranus/Neptune conjunction of the era.

Theodor Landscheidt's work has inspired both professional and citizen scientists. For example, Carl Smith (2007)[7] while researching Landscheidt's work produced an AM graph using the JPL ephemeris. This graph for the first time showed the detailed perturbations of solar AM that also coincide with past solar slowdowns along with the disordered solar path about the SSB. Carl Smith passed away in 2009 and to our knowledge was probably not aware of the hidden detail that was contained in his work, the perturbed Angular Momentum curve holding the clue.

The perturbed curves on the solar AM graph correspond with a solar torque perturbations which also alter the normal balanced solar path around the SSB. The solar velocity is also perturbed on a 172 year cycle (average) and a greater diversion between the orbital AM of the Sun and planets is observed. It is proposed through a spin orbit coupling mechanism resulting in varying solar equatorial rotation rate, the solar dynamo is reduced during these 172 year intervals.

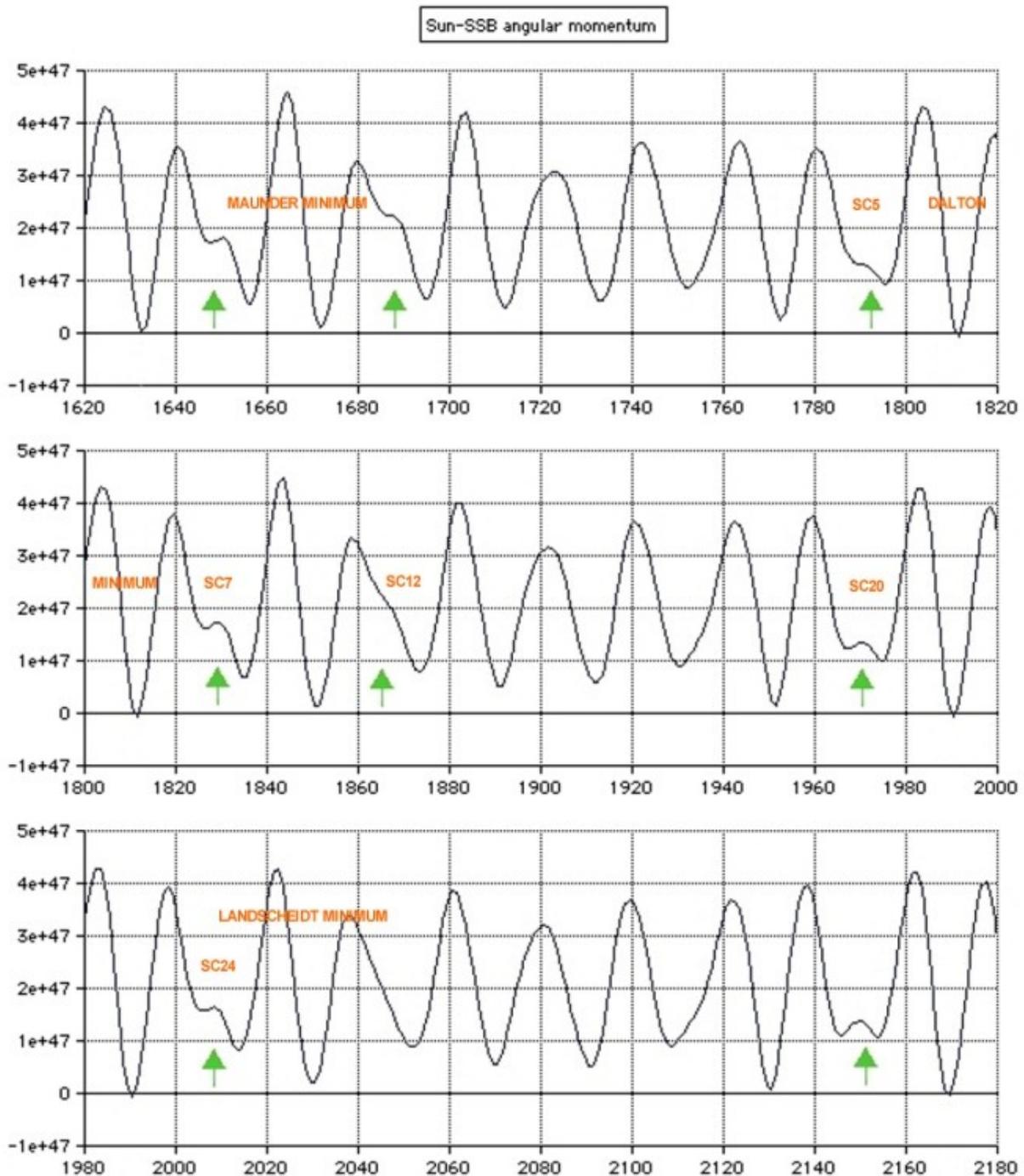

**Figure 1.** Detailed solar AM showing perturbations (AMP) at the green arrows. This graph is a modified example of Carl Smith's original findings. The green arrows and solar grand minima have been added. Carl Smith's [7] original graph is displayed  HERE
Note: AM units of measure equate to gram-cm^2/sec

### Discussion – Grand Minima

The AM perturbations shown in Figure1 are the result of the extra AM from the Uranus/Neptune conjunction. The timing in relation to the Jupiter/Saturn opposition provides the different perturbation shapes that can be measured via the relevant planet angles and categorised into two groups. Perturbations occurring on the down slope "Type A" and those on the up slope "Type B" (down slope = right hand side of peak).

Type B occurrences coincide with weaker solar slowdowns and are more common before 1000AD and the Medieval Warm Period (MWP). See Figure 5.

Almost all perturbations throughout the past 6000 years coincide with solar downturns which vary in intensity.

During the past 750 years strong Type A perturbations are evident on the AM graph. Coinciding with the strong Type A multiple appearances is one of the greatest sustained periods of grand minima of the Holocene (Little Ice Age 1300-1870 approx). The Type A perturbations all have the same planetary configuration but with slightly differing planet angles.

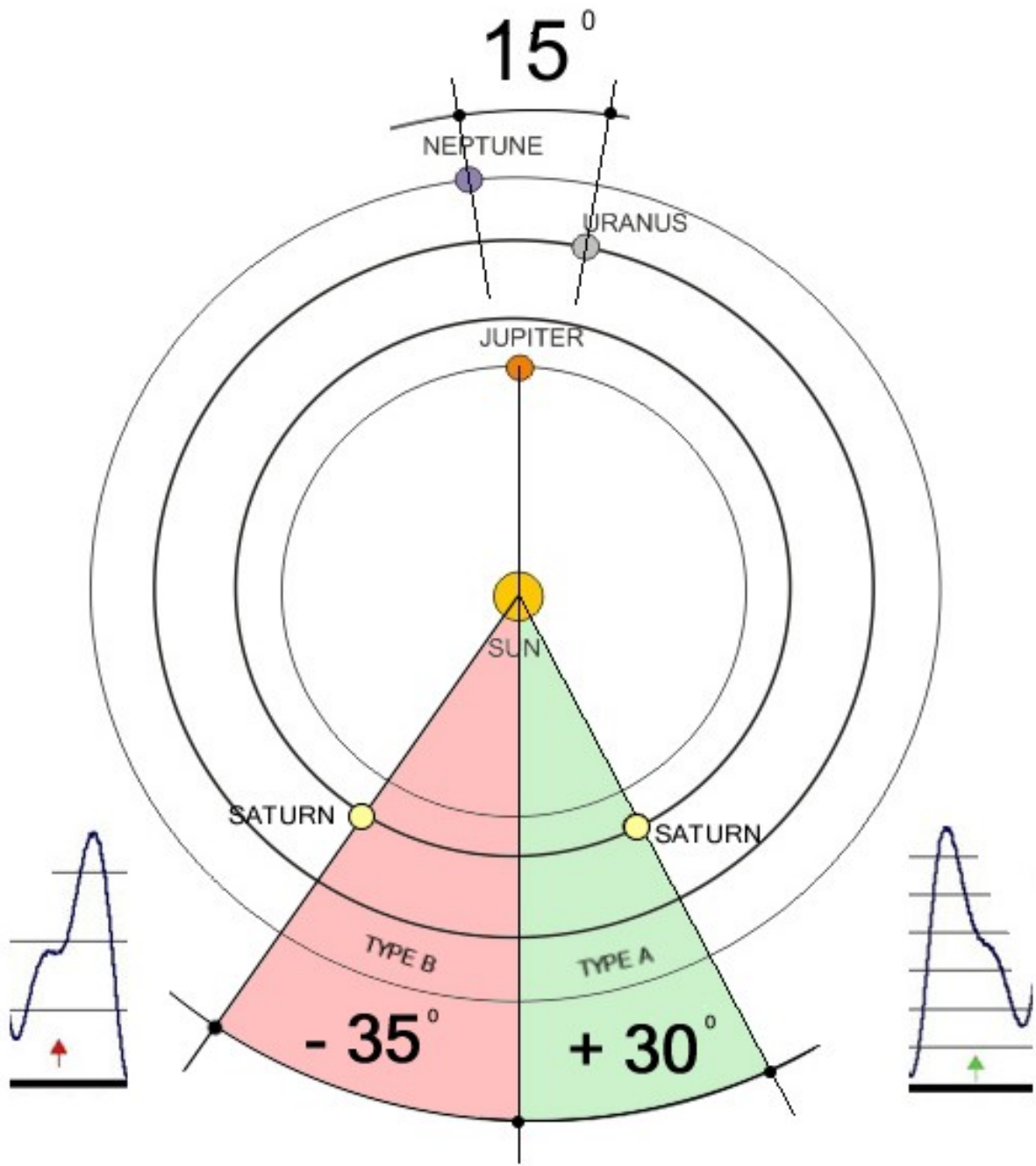

**Figure 2.** Typical planet positions demonstrating strong Type A&B perturbations. The Type A example taken from near the centre of the Sporer Minimum (1472). Type B events coinciding with less reduction of solar activity compared with Type A events of similar angle (reverse).

Type A perturbations have a positive Saturn angle, the higher the Saturn angle the higher the perturbation height on the AM graph. Type B perturbations always display a negative Saturn angle. During each conjunction of Uranus & Neptune, Jupiter & Saturn have multiple oppositions. Depending on the planet positions of the era this can result in normally 3-4 Angular Momentum Perturbations (AMP) events for each Uranus/Neptune conjunction. The midpoint of these AMP groups displays a 172 year average spacing. The Sporer Minimum (1400-1600 approx) has 2 strong AMP events, and 2 medium AMP events which coincide with one of the longest and deepest grand minima of the Holocene.

Type A AMP events have a major impact on the inner loop trajectory of the Sun in its orbit around the SSB. The Sun normally follows two distinct loops around the SSB with each loop lasting approximately ten years. A shallow inner loop is evident when Jupiter & Saturn are in opposition and a much wider loop when Jupiter & Saturn are in conjunction. During strong Type A AMP events the inner loop path is greatly extended pushing the Sun out of its normal balanced trefoil pattern. The normal trefoil pattern returns the Sun to near the SSB on the inner loop path. The AMP event shows the path greatly extended from the SSB. Type B AMP events affect the outer loop of the Sun's path, which has the effect of reducing the distance travelled away from the SSB. Accompanying the AMP events is usually a zero crossing (solar AM goes below zero) where Saturn, Uranus & Neptune are in conjunction with Jupiter opposing. This places the Sun right on the SSB, or indeed on the other side of the SSB. The zero crossings can occur either side of an AMP event.

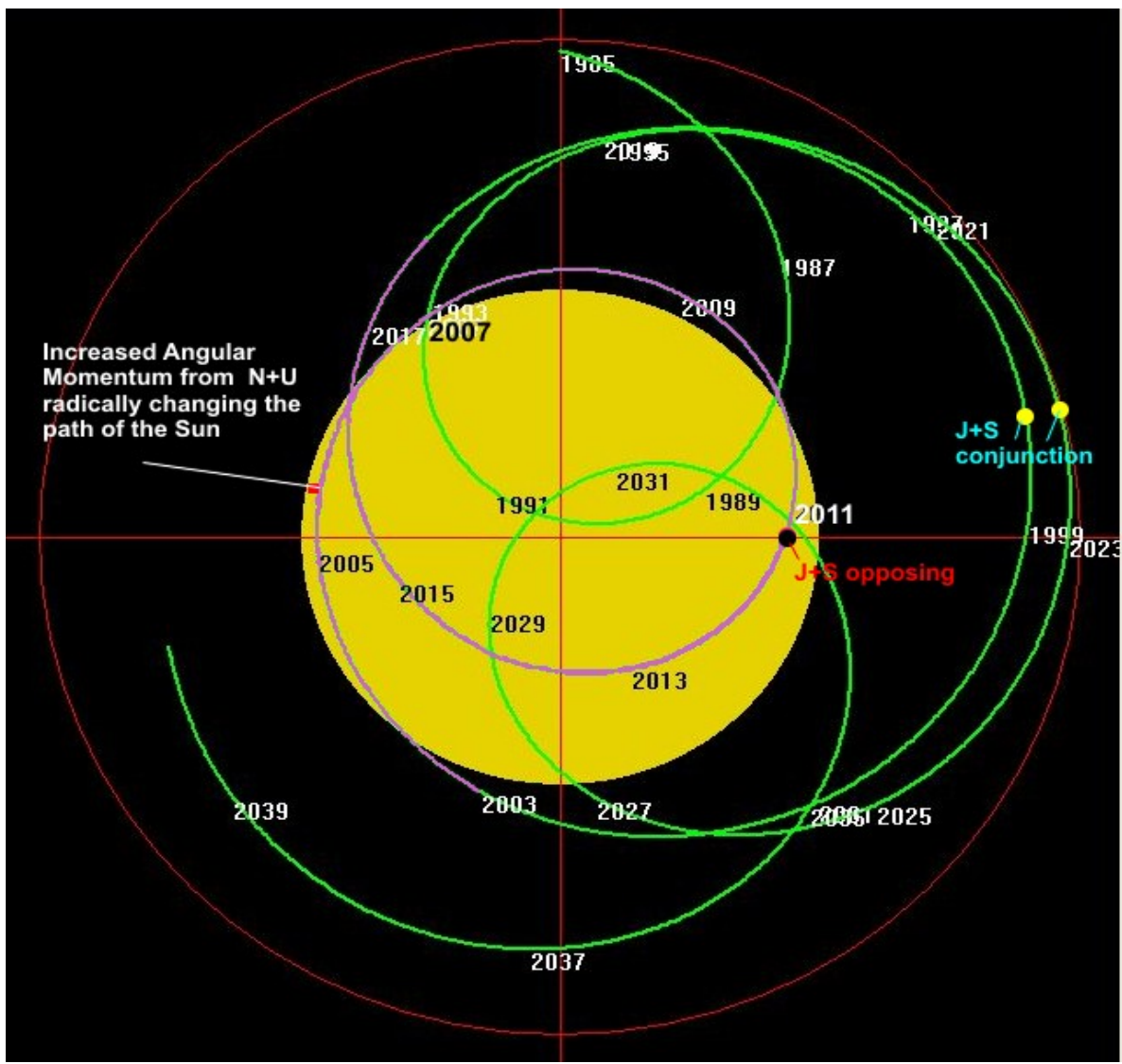

**Figure 3.** The path of the Sun showing the two distinct loops around the SSB (centre point). The extended inner loop beginning around 2005 coinciding with a reasonably strong Type A event. Solar cycle 24 is predicted to be the first grand minimum cycle. The sunspot records since 1650 suggest that 2 solar cycles can be affected by strong Type A events, there is speculation that the Hale cycle is interrupted and follows a 22 year period. An animated movie of the path taken can be viewed at
http://www.landscheidt.info/images/sim.swf

The Carbon 14 (14C) record for the Holocene is a reliable solar proxy record. Recent Beryllium 10 (10Be) isotope records derived from ice cores by Steinhilber, Beer, Frolich (2009)[9] confirm the accuracy of the 14C record. During the 11,500 years of the Holocene a regular pattern of solar downturns can be observed which vary in intensity. The 14C records used in this report are originally from the INTCAL98 (Stuiver et al.,1998b)[10] study and further extended by Solanki *et al* (2004)[8] and Usoskin, Solanki & Kovaltsov (2007)[11]. The 14C values are compared with the AMP group centre values as well as each individual AMP event. The results show a strong correlation with each individual AMP timing and strength. Each AMP group centre has the relevant planet angles recorded including the orientation of Uranus/Neptune (which planet is leading)

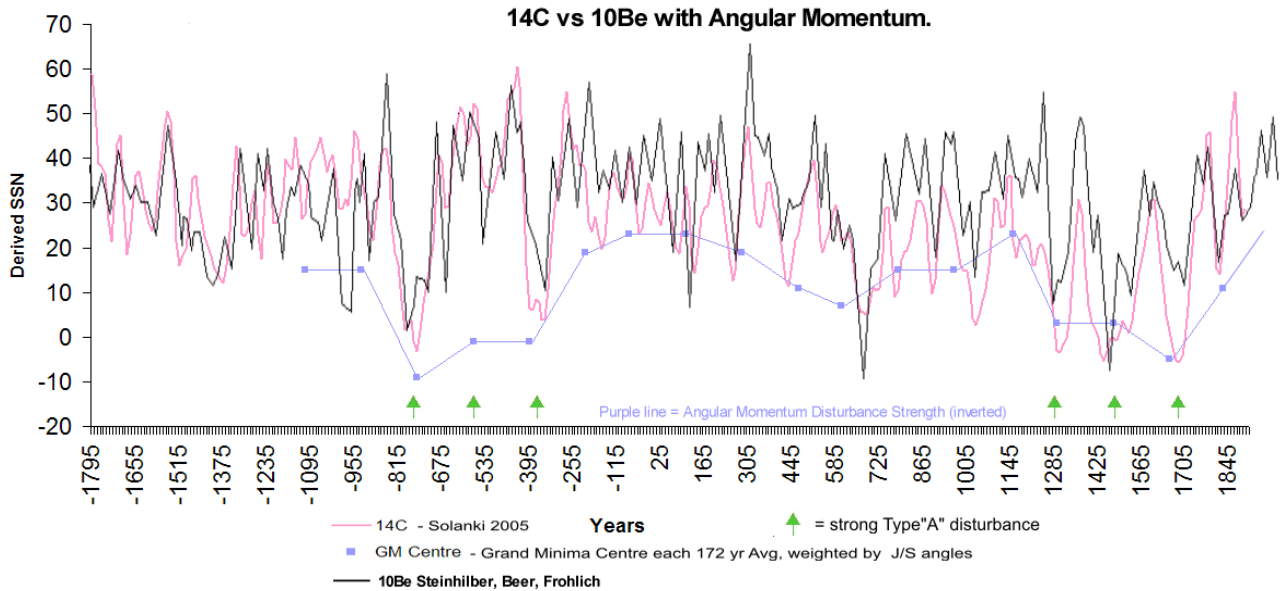

**Figure 4.** Comparison of 14C and 10Be isotope records. The purple line is a representation of the AMP group strength (individual AMP events are summed to form group totals, details follow in later section) with each point representing the AMP centre. A full size image can be viewed at http://www.landscheidt.info/images/solanki_sharp.png  the isotope plots are a graphical comparison of Solanki *et al*(2004)[8] & Steinhilber *et al* (2009)[9]

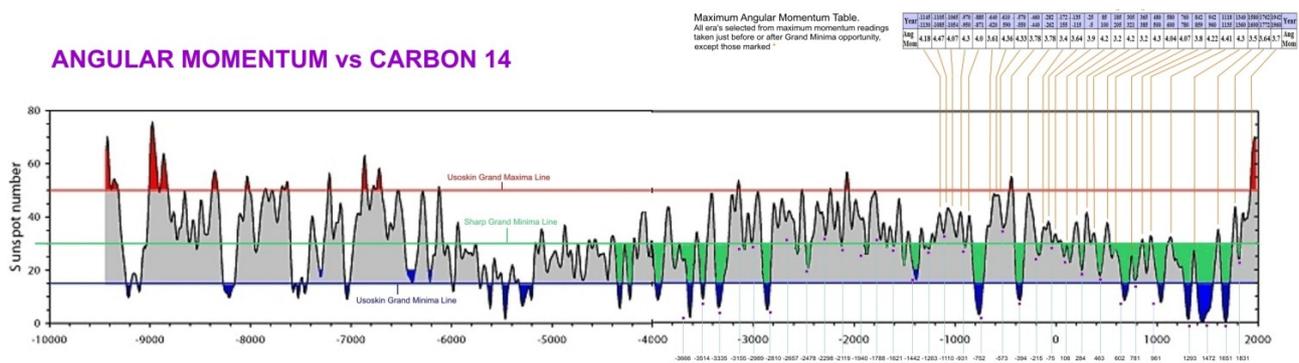

**Figure 5.** Graphical representation modified from the Usoskin *et al*. (2007)[11] solar proxy report . Usoskin determining grand minima occur under 15SSN (derived SSN figures) which isolates Dalton Minimum type events. By raising the bar (green line) the repeating pattern of grand minima is observed. The middle AMP event per group shown, the total of the AMP group determining the overall solar downturn of the era. Figure 7 shows each individual AMP event making up each AMP group, the strength of each disturbance coinciding with the relevant planet angles. Type A events with a J/S angle around +30 deg and N/U angle of 15 deg providing the strongest downturns, Type B&A events with a J/S angle near zero deg or 180 deg providing the weakest downturn. A full size image can be viewed at http://www.landscheidt.info/images/c14nujs1.jpg

| Year | -3666 | -3514 | -3335 | -3155 | -2989 | -2810 | -2657 | -2478 | -2298 | -2119 | -1940 |
|---|---|---|---|---|---|---|---|---|---|---|---|
| Pos | N/L | N/L | N/L | N/L | N/L | N/L | N/L | N/L | N/L | U/L | U/L |
| N/U | 8 | 55 | 27 | 10 | 23 | 2 | 45 | 38 | 7 | 5 | 14 |
| J/S | -55 | +32 | +32 | +40 | -78 | -95 | +2 | +10 | 0 | -12 | -13 |

| Year | -1788 | -1621 | -1442 | -1263 | -1110 | -931 | -752 | -573 | -394 | -215 | -75 | 106 |
|---|---|---|---|---|---|---|---|---|---|---|---|---|
| Pos | N/L | N/L | N/L | Draw | N/L | N/L | N/L | Draw | U/L | U/L | N/L | N/L |
| N/U | 13 | 22 | 13 | 2 | 30 | 30 | 16 | 0 | 17 | 42 | 30 | 13 |
| J/S | +85 | -35 | -34 | -45 | +58 | +62 | +35 | +18 | +9 | 0 | 0 | -10 |

| Year | 284 | 463 | 602 | 781 | 961 | 1153 | 1293 | 1472 | 1651 | 1831 | 2010 |
|---|---|---|---|---|---|---|---|---|---|---|---|
| Pos | U/L | U/L | N/L | N/L | N/L | U/L | N/L | N/L | U/L | U/L | U/L |
| N/U | 3 | 35 | 40 | 30 | 15 | 30 | 48 | 15 | 5 | 20 | 40 |
| J/S | -24 | -36 | -30 | -35 | -65 | +53 | +55 | +30 | +25 | +25 | +25 |

**Figure 6.** Planet angle tables referring to Fig 5. displaying the middle AMP event per 172 year average period. Pos= relative Uranus/Neptune position ie Neptune leading Uranus etc. N/U= angle measured between Neptune & Uranus. J/S= angle measured away from Jupiter/Saturn opposition.

The planetary angles taken at the AMP events providing a statistical measure that can be compared with the AMP events and Figure 1. The depth of the solar downturns shown on the 14C graphs coinciding with the planet angles and observed AMP events, deep troughs aligning with strong Type A events and sustained periods of strong solar activity aligning with periods of weaker Type B events. Type A events can also be weak depending on the planet angles. Type B events occurring before 1000AD are a result of the changing planet angles that move slowly over long periods of time. The overall background shape of the 14C graph coinciding with the occurrence and strength of Type A & B events. During times of multiple sustained Type B events each 172 year cycle can carry more than 3 events as observed in Figure 7.

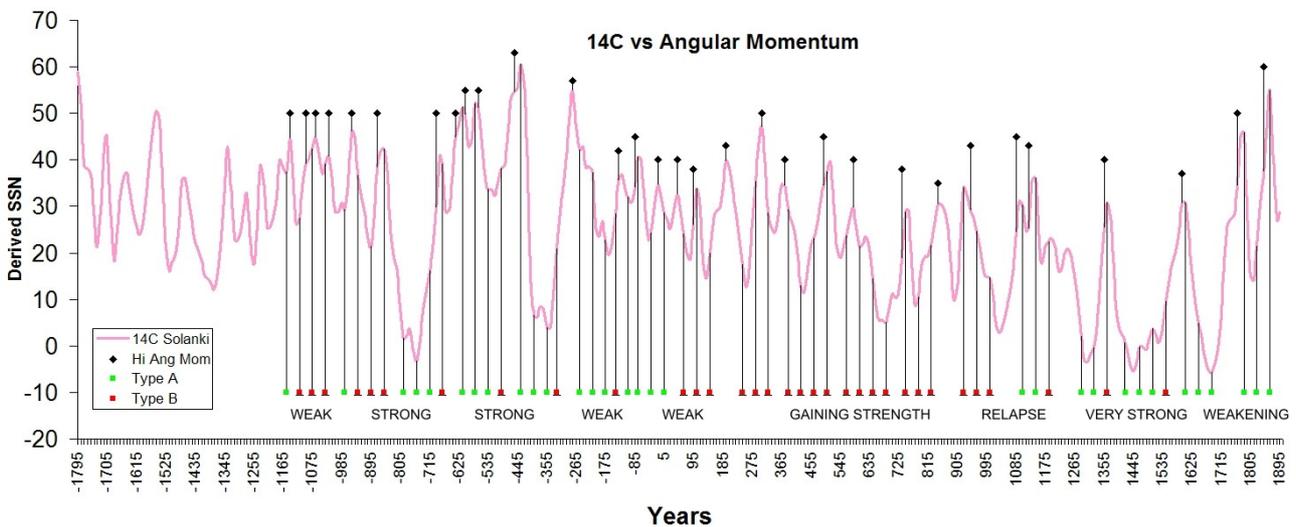

**Figure 7.** Individual AMP events plotted directly onto the Solanki data[8]. As each conjunction of Uranus & Neptune has a varying Jupiter/Saturn position, the strength of each individual AMP event along with the AMP group total changes over the millennia. This can be represented as the AM power curve of the Holocene. A full size image can be viewed at
http://www.landscheidt.info/images/solanki_sharp_detail.jpg  spreadsheet with original Solanki data[8] with AMP events available at
http://www.landscheidt.info/images/solanki_sharp.xls

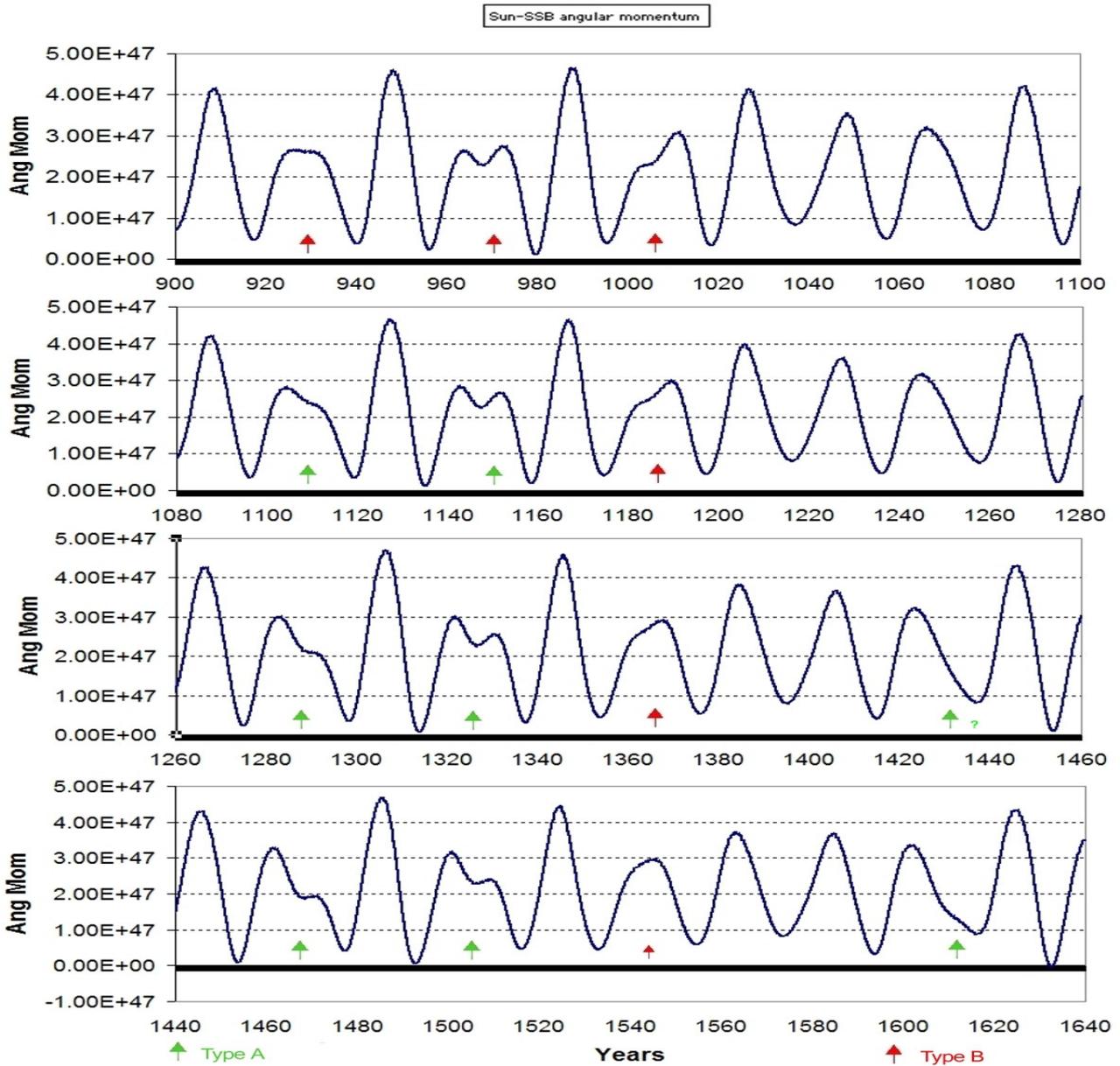

**Figure 8.** AM graph depicting the period from 900AD – 1640AD. The change in Type A dominance shown after 1100AD. The MWP possibly being the only period during the Holocene not to be affected by the recurring 172 year AMP pattern. This is a time of transition moving from very weak Type B occurrence to a strengthening Type A dominance. The AMP events at 930 & 970 are midway between Type A & B.  Note: AM units of measure equate to gram-cm^2/sec

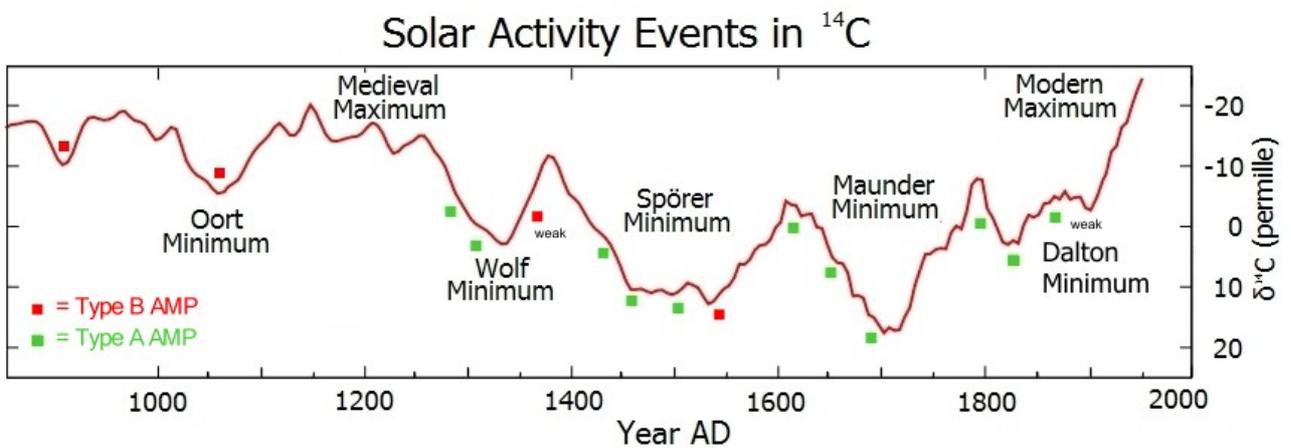

**Figure 9.** Carbon 14 graph (Wikipedia) with the AMP events overlaid. The Sporer Minimum displaying the longest period of solar inactivity coinciding with 3 strong Type A events and 1 Type B event. The last two AMP events during the Maunder Minimum being stronger than the initial AMP event. See Figure1&8.

**Determining AMP Strength**

The purple line shown on Figure 4 is a representation of the AMP strength of the era which follows the general trend of solar activity. The method used is a preliminary method using visual observation of each disturbance of the graph period. Figure 1 displays the different types of AMP events that cycle in groups every 172 years (average). These disturbances always line up with periods of solar downturn.

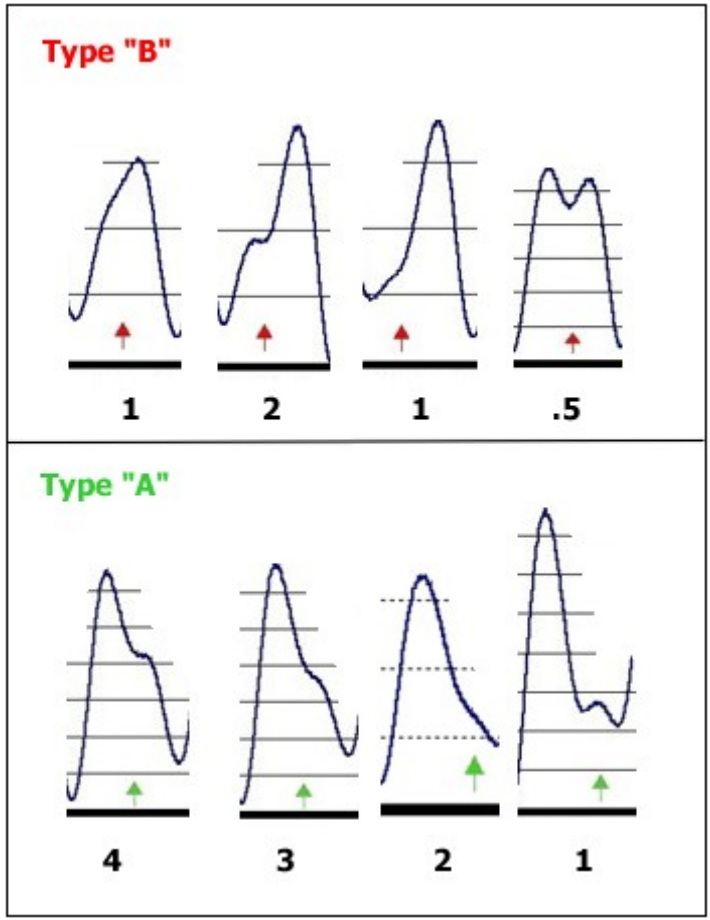

| Year | Count | Conv |
|------|-------|------|
| 1830 | 5 | 11 |
| 1651 | 9 | -5 |
| 1472 | 7 | 3 |
| 1293 | 7 | 3 |
| 1153 | 2 | 23 |
| 961 | 4 | 15 |
| 781 | 4 | 15 |
| 602 | 6 | 7 |
| 463 | 5 | 11 |
| 284 | 3 | 19 |
| 106 | 2 | 23 |
| -75 | 2 | 23 |
| -215 | 3 | 19 |
| -394 | 8 | -1 |
| -573 | 8 | -1 |
| -752 | 10 | -9 |
| -931 | 4 | 15 |
| -1110 | 4 | 15 |

| Count | 2 | 3 | 4 | 5 | 6 | 7 | 8 | 9 | 10 |
|-------|---|---|---|---|---|---|---|---|----|
| Conversion | 23 | 19 | 15 | 11 | 7 | 3 | -1 | -5 | -9 |

**Figure 10.** Charts showing the AMP strength quantification method. Each count unit corresponding with 4 units of Usoskin[11] derived SSN units. Future quantification methods involving accurate planet angles would provide more detail.

The Solanki/Steinhilber[8][9] data shows regular solar downturns that vary in intensity, by observing the shapes of the AMP events that align with these downturns we are able to see a pattern that is repeatable.

There are rare occasions of strong Type A AMP events that do not cause solar activity reduction or perhaps not as low as expected, when not meeting the Wilson Test described in Wilson *et al*. (2008)[12]. This test states that for an AMP event to fully utilise the disturbance, the Jupiter/Saturn opposition or conjunction must happen before cycle maximum (to achieve a cycle with less than 80SSN). This has been tested over the sunspot record but is not available for accurate testing beyond that as the cycle maximum date is not known. 1830 and -530 are probable examples of this phenomena, during 1830 the Jupiter/Saturn conjunction occurred before cycle maximum. Cycle maximum records not available for -530.

By matching the AMP event shapes on the AM graphs with solar downturn strength each disturbance can be quantified. AMP events that align with deep grand minima (on a constant basis throughout history) get the highest score and also show the same shape or perturbation. Each point on the purple line shown on Figure 4 is a total of all AMP events making up that 172 year group.

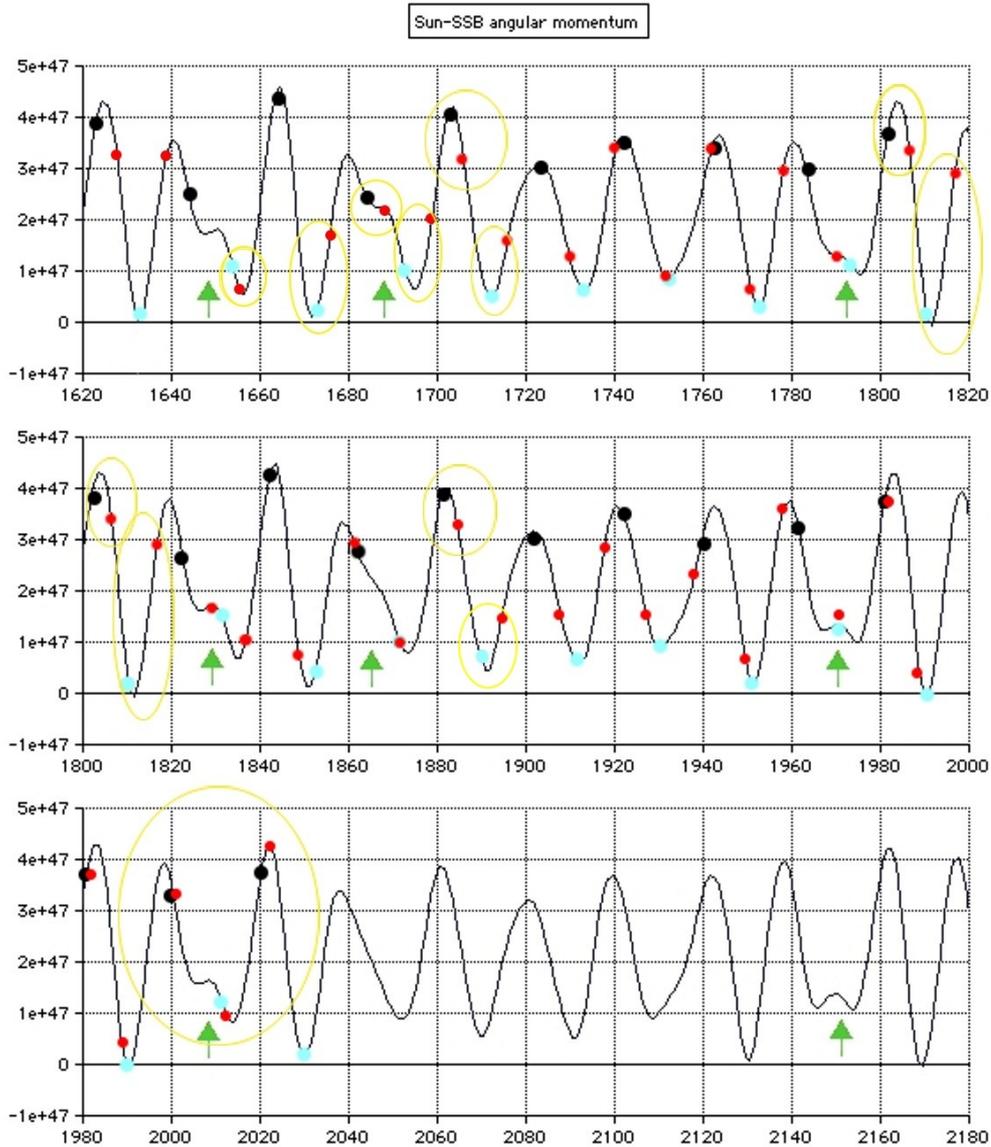

**Figure 11.** Wilson's Test. Black dots are Jupiter/Saturn together, Blue dots Jupiter/Saturn opposed, and Red dots are solar cycle maxima, reduced solar activity occurs when a black or blue dot occurs in between cycle minimum and before cycle maximum. 1830 & 1790 not passing the test. Yellow circles complying with Wilsons law. Note: AM units of measure equate to gram-cm^2/sec

### Discussion – Solar Cycle Modulation

Solar cycle strength depicted by SSN (smoothed sunspot number) follows a recurring wave of power that follows the overall AM values (Figure 14). This AM wave displays a peak at the Uranus/Neptune conjunction and the corresponding trough at the Uranus/Neptune opposition, Scafetta (2009, 2010)[6] also recognised this trend in his work dealing with the celestial origin of climate oscillations. The SSN record of the past 400 years when matched with this wave shows a strong correlation. The AMP events that always occur at different intensities and length near the AM maximum, interrupt the correlation and work as a separate process. The AM sine wave is also seen as a background solar engine that affects cycle modulation (SSN) but not the length of the cycle. Other processes determine the length of each solar cycle.

The Damon Minimum (1856-1913) is sometimes described as a grand minimum and is most likely an example of a low sunspot cycle/s affected by low AM at the time of the Uranus/Neptune opposition and not a true grand minimum. AMP events are not observed during these intervals with only low AM recorded. The AMP event is thought to create a "phase catastrophe" situation which perhaps is responsible for reported monopolar solar pole readings of the Maunder Minimum by Callebaut *et al*, (2007)[10]. The monopolar position is effectively providing the majority of sunspot activity to a single solar hemisphere. The monopolar position being described as a prolonged period where the solar poles have the same polarity as a result of only one pole reversing polarity during the Hale cycle. These events need to be accommodated when comparing the AM modulation versus the sunspot cycle modulation. The monopolar disturbance to the normal Hale cycle could explain the occurrence of twin low cycles paired during grand minima even though the second cycle is not perturbed. The poles require the extra cycle to maintain the normal synchronised state. The key point being that low cycles can be a result of low AM without experiencing "phase catastrophe" conditions. Grand minima occur during the higher part of the AM wave.

The AM graphs show a sine wave of AM modulation (around 10 years), the low points are considered as important as the high points. While a direct mechanical link between AM modulation and solar cycle modulation remains theoretical there is a direct example of some physical connection. The solar velocity around the SSB is absolutely linked to the AM sine wave which provides a roughly decadal acceleration/deceleration phase.

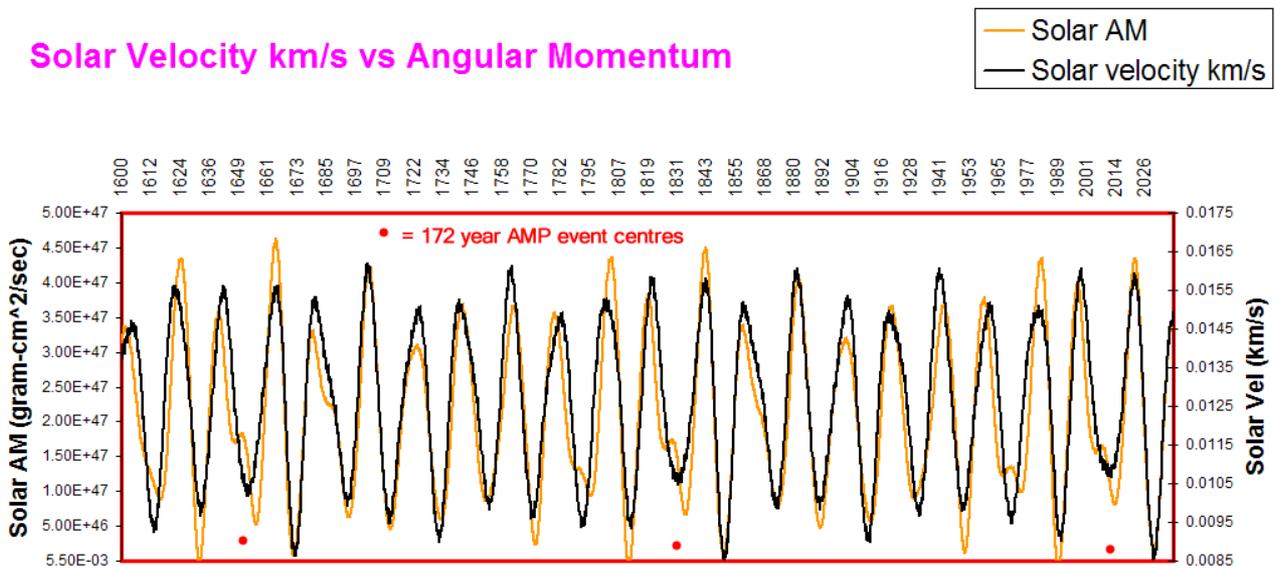

**Figure 12.** Solar AM values matched with solar velocity. The red dots displaying the 172 year centre of the AMP groups. Note the repeating pattern of change in velocity. Data source- Jet Propulsion Laboratory.

To visualise the importance of the low points in relation to the high points of the AM graph, a centre point needs to be determined and the values recorded under that centre point are inverted. This provides a true reading of the AM strength. For the past 400 years an AM reading of 2E+47 (gram-cm^2/sec) was used from Carl Smith's data as the centre point.

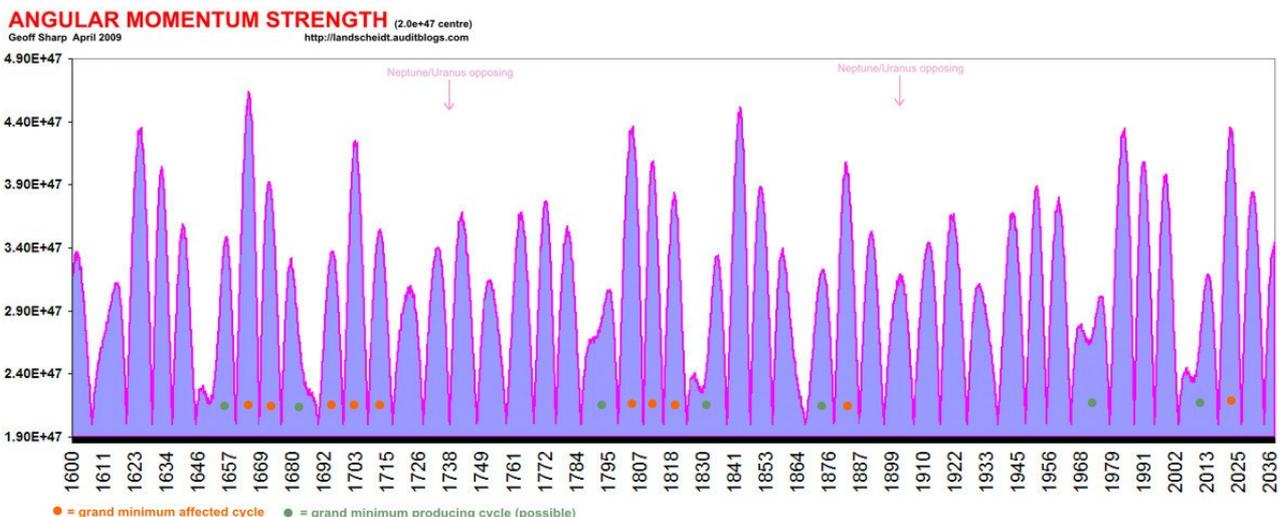

**Figure 13.** Manipulated AM graph using Carl Smith's original data with all points below 2E+47 inverted.

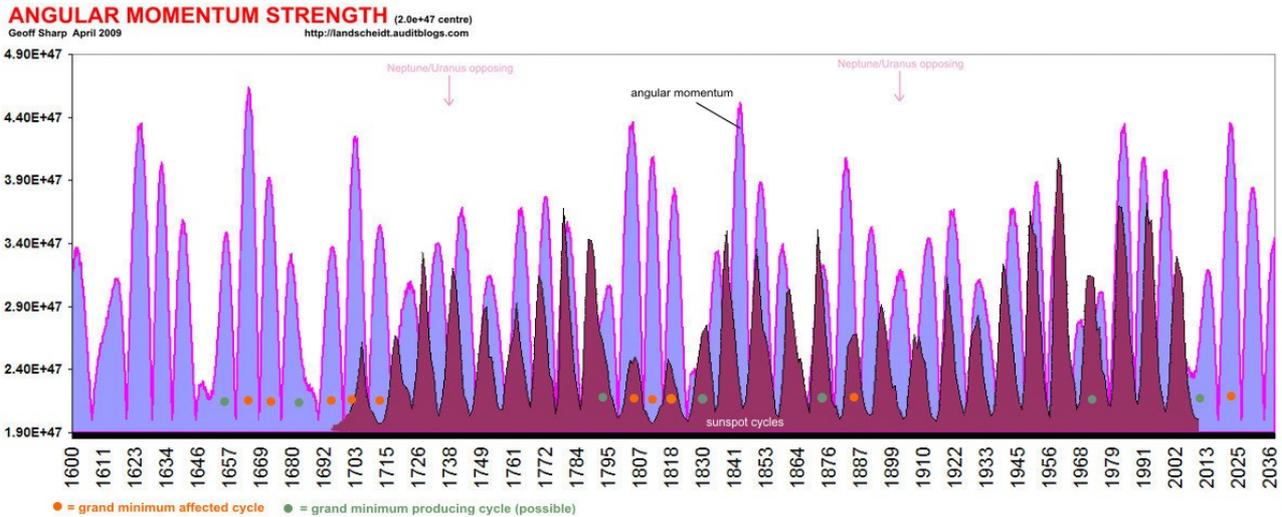

**Figure 14.** Inverted AM values with the SIDC SSN values overlaid graphically.

**Angular Momentum Data**

The original Carl Smith AM[7] data is referenced and validated by independent analysis carried out by G.E. Pease using JPL coordinates and the following standard formula:

$$L = M\sqrt{(y\dot{z} - z\dot{y})^2 + (z\dot{x} - x\dot{z})^2 + (x\dot{y} - y\dot{x})^2}$$

M is the Mass of the Sun in kilograms. x, y, z, xdot, ydot, zdot must be converted from kilometres and km/sec to metres and metres/sec to get m^2ks units. The equation yields the absolute value of L, using the instantaneous cross products of the body's position and velocity vectors.

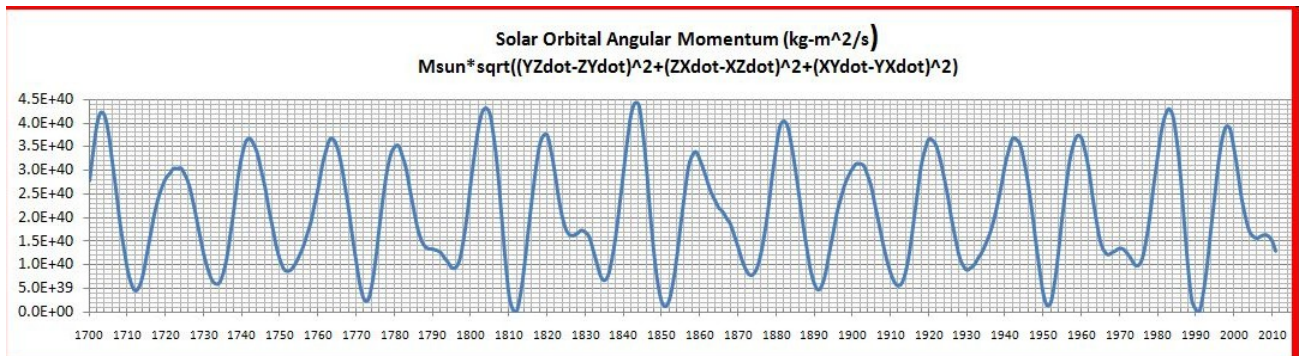

**Figure 15.** G.E. Pease AM graph depicting the same AMP events and timing as the Smith AM graph.

**Proposed Mechanical Link Via Spin Orbit Coupling**

A spin orbit coupling mechanism has been discussed by Wilson *et al.* (2008)[12] which translates into a varying solar equatorial rotation speed, enabling changes to the solar dynamo and the meridional flows. Solar equatorial rotation rate changes have been recorded by Javaraiah (2003)[3] based on sunspot movement records.

Total Angular Momentum is the combination of orbital AM and spin AM. The laws of AM conservation allow a "trade off" between orbital AM and spin AM. Each body in the solar system has its own orbital AM that can be calculated using the standard formula:

$$L = M\sqrt{(y\dot{z} - z\dot{y})^2 + (z\dot{x} - x\dot{z})^2 + (x\dot{y} - y\dot{x})^2}$$

If there is a discrepancy between solar orbital AM and planet/body orbital AM the laws of AM conservation would allow changes to a body's spin AM. This could result in a varying solar equatorial rotation rate.

To perform this task all the solar system planets and the asteroids Ceres, Juno, Vesta, Pallas, Eugenia, Siwa & Chiron have been included to arrive at a total planet AM. Data coordinates were taken from the JPL DE405 ephemeris.

To compare planet AM with solar AM the inertial frame should be the same. The JPL DE405 heliocentric planetary coordinates are referred to the solar system barycentric inertial frame rather than the heliocentric inertial frame, which required a transformation of our computed planetary angular momenta to the heliocentric inertial frame. The planet AM was calculated using heliocentric coordinates, the Sun was calculated using the SSB as the axis point (barycentric), then the solar AM is subtracted from the planet AM to achieve the same inertial frame.

The following graph displays a divergence between solar and planet orbital AM. A future study will be performed with G.E. Pease further outlining this procedure. Extending this graph back over the whole LIA may prove interesting, possibly showing us another method of identifying solar slow down by studying the planetary AM and its relationship with solar AM. Some big questions remain.

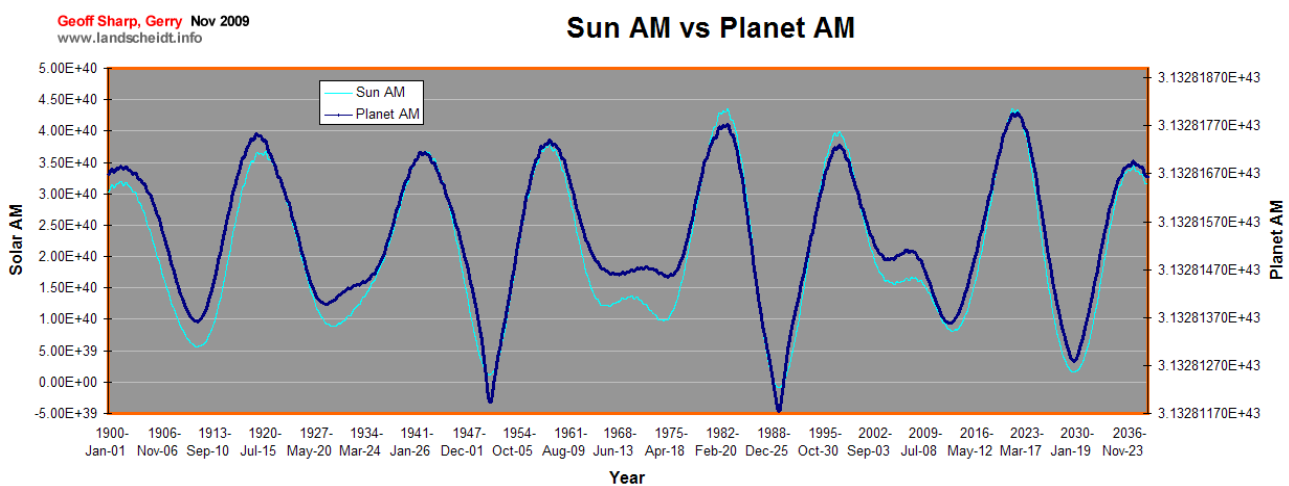

**Figure 16.** Differences in orbital AM become clear when viewed in the correct inertial frame. Note the near constancy (smaller range of deviation from top to bottom) of the Planet AM after subtracting the solar AM. The Planet AM is now constant to six significant figures, whereas it was only constant to three significant figures before the inertial frame transformation. We believe the residual variations in the seventh and higher significant digits may be the result of spin orbit coupling between the planetary orbits and solar rotation.

**Conclusions and Predictions**

The correlation of the inverted AM values with the existing sunspot record, along with the quantification of AMP events provides a platform for future sunspot prediction out to 3000AD. Cycles 24 & 25 are predicted to be less than 50SSN using the Layman's Sunspot Count (based on the SIDC values but ignoring specks rated lower than 23 pixels). Solar cycle 20 was the first stage of the current AMP group which failed to generate a full grand minimum because of the very weak AMP event caused by the late timing of the Uranus/Neptune conjunction and the failed Wilson's Test. The current AMP group does not display a third event which is extremely rare and hence will allow a modest recovery during solar cycle 26. Looking out further, the next 1000 years do not show any major chances for deep grand minima which should provide stable conditions for future generations, not withstanding the possible entry into the next ice age.

Solar cycle 24 will need to reach maximum after March 2011 to comply with the Wilsons Test.

Further information on the Layman's Sunspot Count can be found at:

http://www.landscheidt.info/?q=node/50

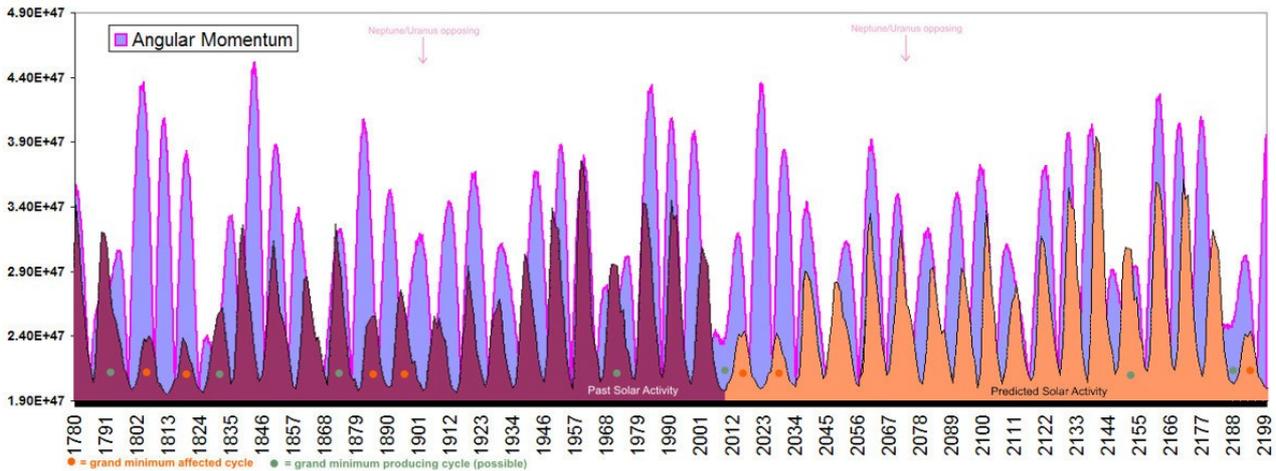

**Figure 17.** 200 year solar cycle prediction. The trough in the sunspot record coinciding with the opposition of Uranus & Neptune in past records and is expected to remain the same. Solar proxy records do not show high solar activity during times of low AM (Figure5&7). The length of each future solar cycle remains unknown, predicted solar cycles should be taken as a guide with the overall trend being important.

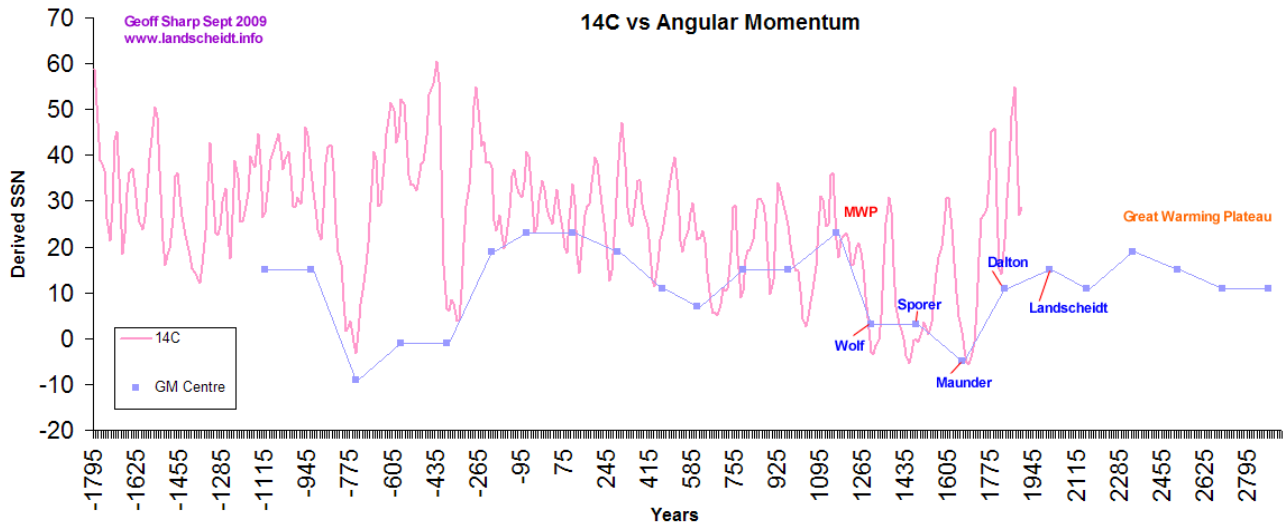

**Figure 18.** Earth's future climate based on the projected solar activity using the same method as described in the chapter headed "Determining AMP Strength". Deep grand minima are not expected.

## Acknowledgements

Special thanks go to G.E. Pease for confirming Carl Smith's AM data and for the advice and input on inertial frames.

Also many thanks to Nicola Scafetta for providing advice and initial peer review.

The SIM diagram (Figure 3) produced on software made available by Carsten Arnholm.

**Extra AM Data**

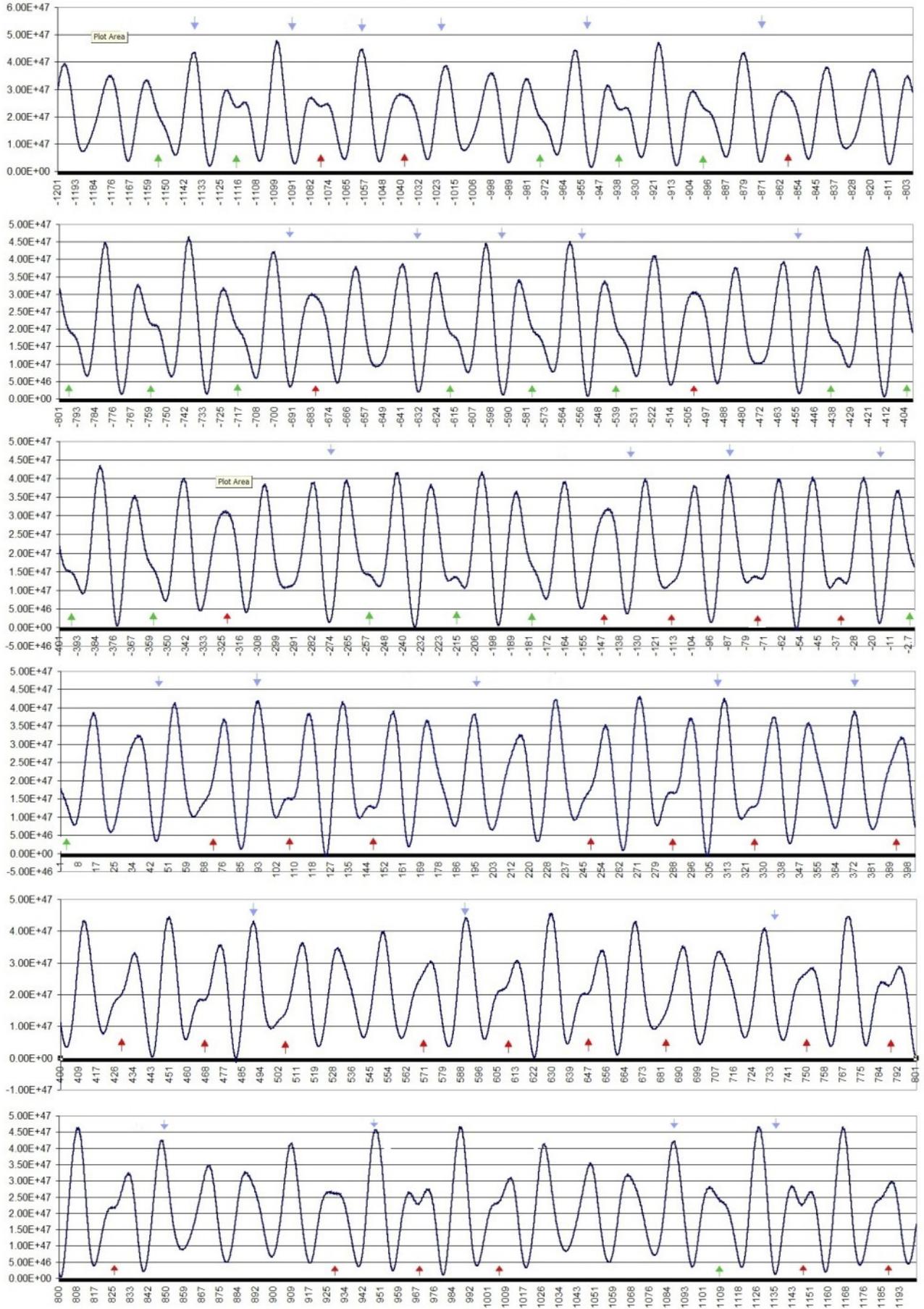